\parindent=0pt

 \baselineskip=14pt

\centerline{\bf  On Effective Constraints for the Riemann-Lanczos System of Equations.}

\

\

\centerline {S. Brian
Edgar}

\

\centerline{Department of Mathematics, Link\"{o}pings universitet,
 Link\"{o}ping, Sweden S-581 83.}

 \centerline{E-mail: \ bredg@mai.liu.se}

\

\

{\bf Abstract}  

There have been conflicting points of view concerning the Riemann--Lanczos problem in 3 and 4
dimensions.  Using direct differentiation on the defining  partial differential equations, Massa and Pagani (in
4 dimensions) and Edgar (in dimensions $n \ge 3$) have argued that there are effective constraints so that not 
all Riemann tensors can have Lanczos potentials; using Cartan's criteria of integrability of ideals of
differential forms Bampi and Caviglia have argued that there are no such constraints in dimensions $n\le 4$, and
that, in these dimensions,  all Riemann tensors can have Lanczos potentials.  In this paper we give a simple
direct derivation of a constraint equation, confirm explicitly that known exact solutions of the Riemann-Lanczos
problem satisfy it, and argue that the Bampi and Caviglia conclusion must therefore be flawed. In support of
this, we refer to the recent work of Dolan and Gerber on the three dimensional problem; by a method closely
related to that of Bampi and Caviglia, they have found an 'internal identity' which we demonstrate is precisely 
the three dimensional version of the effective constraint originally found by Massa and Pagani, and Edgar.

\

\vfill\eject

{\bf 1. Introduction.}

In two recent papers Dolan and Gerber [1,2] have revisited the {Riemann--Lanczos problem} [3],
i.e., whether a Riemann tensor $R_{abcd}$ can
be generated from a 3-index tensor potential $H_{abc}$,
$$R_{abcd} = \  2 H_{ab[c;d]}
+2H_{cd[a;b]} \eqno(1)$$
where the potential $H_{abc} $ satisfies 
$$H_{abc}=H_{[ab]c} \qquad \qquad
H_{[abc]}=0.
 \eqno(2)$$
In the literature\footnote{${}^{\dag}$}{We follow the conventions of [4]. Since there will be
expressions which involve sums of terms with respectively  the Ricci tensor and the square of
Ricci tensor, a consistent convention is essential. In particular it should be noted that in [5,6,7] a
different convention was used  for the Ricci tensor
and so there are some sign differences between some equations in this paper and their counterparts
 in [5,6,7].}, there are two apparently conflicting answers to this problem in four dimensions:
Massa and Pagani [5] have argued that: 

{\it ... for the class of spacetime metrics satisfying $R_{ab}=\lambda g_{ab}$  one of the
integrability conditions of the system (1) takes the form $R^2-2R_{abcd}R^{abcd}=0$, i.e., it
imposes a restriction on the geometry itself }

whereas Bampi and Caviglia [8] have argued that:

{\it..., in the four-dimensional case no integrability
condition is required. In other words, looking at the class of singular solutions allows a 
potential to exist
without any restriction .... on the geometric structure of the underlying
Riemann manifold  }

These two papers were written at the same time, and neither refers to the other.  Of course this uncertainty would
be dispelled if there were explicit examples of Lanczos potentials in reasonably general spaces; however, the only
known examples are a few in flat [3,9] and conformally flat spaces [9] and a few very special spacetimes in
[1,2].

Dolan and Gerber [1,2] rely  on the results and to some extent on the method from [8], while  stating that the
paper of Massa and Pagani in [5], and the subsequent work by Edgar in [6,7], 

{\it ... uses a totally
different approach ... and is not
applicable here}.

We agree with Massa and Pagani [5] that if the representation (1) exists for arbitrary
Riemann tensors then it would have very significant implications for
both the mathematical  viewpoint and physical applications of general
relativity.  Therefore it is important to determine
whether the representation (1) exists for all spaces (as  argued in [8,1,2]), or whether it only
exists for a  restricted number of spaces (as argued in [5,6,7]). If the latter case, it would be interesting
to know whether there exist  other effective constraints.  Moreover, we believe that the work in
[5,6,7] is very applicable to the work in [8,9] and [1,2], and a fuller understanding of the links between these two
lines of investigation  should clarify the apparent contradiction.

So, in this paper, we shall first demonstrate in a very simple unambiguous
manner that (1) implies an integrability condition which is
an effective constraint for dimensions $n > 2$ and therefore limits the class of spaces which can
permit a Lanczos potential with properties (2) via (1). Moreover, we will show explicitly,
that the Lanczos potentials of those very special spaces found in [1,2,3,9] satisfy this restriction. 
Furthermore, we demonstrate that the nontrivial 'internal identity' found  recently  by Dolan and
Gerber in the three dimensional problem [1] is precisely the effective constraint found in [5,6,7], and  we
argue that the Bampi-Caviglia analysis [8] is therefore flawed. We propose that the
Janet-Riquier approach used in [1] should be applied to other dimensions. 

\

{\bf 2. Effective Constraints on Riemann--Lanczos system.} 

\underbar{\it Riemann-candidate--Lanczos problem in
$n$-dimensional spaces}

We consider first the more general  problem in
$n$-dimensional spaces  of whether any  {\it   Riemann-candidate tensor, $\hat R_{abcd}$}
--- a tensor  having the algebraic index symmetries of the Riemann tensor --- can be generated
from a potential $H_{abc}$ by
$$\hat R_{abcd} = \  2 H_{ab[c;d]}
+2H_{cd[a;b]}   \eqno(3)$$
where the potential $H_{abc}$ satisfies (2).

\underbar{Flat space}

It is trivial to show that in flat space, although at the first derivative level we can only eliminate one of
the  potential terms,
$$\hat R_{ab[cd;e]} = \  
2H_{[cd}{}^{[a;b]}{}_{;e]}   \eqno(4)$$
by taking another derivative we can obtain
$$\hat R^{[ab}{}_{[de;f]}{}^{c]}=0 . 
\eqno(5)$$
It is important to check whether (5) is an {\it effective} constraint 
for arbitrary
$\hat R_{abcd}$ --- in the sense that not all Riemann-candidates $\hat R_{abcd}$ satisfy (5) --- or whether there
are situations when the lefthand side is identically zero; it is obvious that the left hand side is identically
zero for dimensions
$n=2$, but in all other dimensions it is an effective constraint.\footnote{${}^{\ddag}$}{This is an important
point, and the effectiveness of the constraint should not  be taken for granted. (In the
Weyl-candidate--Lanczos problem [9,10,11,12,13] a similar calculation gives an analogous equation which turns
out to be trivially satisfied in $4$ and $5$ dimensions [14,15]; this was because any
\underbar{trace-free} tensor
$A^{[abc]}{}_{[def]}$ is identically zero in dimensions $n\le 5$ [16,17].) To prove the effectiveness of this
constraint we can simply choose a local Cartesian coordinate system ${x^i}, i=1,...,n$ for $n > 2$ and
construct a simple counterexample, e.g., $R_{1312} = R_{1213} = \sin x^3  \sin x^2 $, all other Riemann
tensor components zero. } 

Therefore the representation (3), in flat space ($n > 2$), is  valid only for the
subset of Riemann-candidates which satisfy (5). This is an important example which illustrates two principles: it
shows that a particular constraint on the geometry of the space, i.e., putting the curvature tensor
$R_{abcd}=0$, can imply a specific constraint on the Riemann-candidate $\hat R_{abcd}$; on the other hand,
although the constraint is quite restrictive, yet it does  permit a significant class of exceptions. For
instance, a set of Riemann candidates which satisfy  the 'flat space Bianchi equations', $\hat R_{ab[cd;e]}=0$;
in this case
$H_{abc}=(h_{bc;a}- h_{ac;b})$ and $\hat R_{abcd} $ has the form of the Riemann tensor in the linearised theory
[3,9]. Of course, this does not say anything \underbar{directly} about the Riemann--Lanczos problem, since in
flat space the constraint is trivially satisfied. However we believe that the curved space analogue of this
constraint (5) is the crucial equation in the Riemann-candidate--Lanczos problem and the Riemann--Lanczos
problem.

\underbar{Curved Space.}

In general curved spaces we can carry out on (3) the same differentiation steps as led to (5), but
this time the right hand side of equation (5) becomes  complicated. However, we can
easily find a much simpler subset, with significant properties.
Noting that $$\hat R_{ab} \equiv \hat R^i{}_{aib} = 2 H^i{}_{a[i:b]}+2
H^i{}_{b[i:a]}\qquad \qquad \hbox{and} \qquad \qquad \hat R \equiv \hat R^i{}_i = 4
H^{ij}{}_{i:j}$$
we find that, after some rearranging of (3)\footnote{${}^{\dag}$}{The left hand side of (7) is equivalent to
$3\hat R^{[ab}{}_{[ab;c]}{}^{c]}$ for $n > 2$; we  begin instead with $\hat R^{;a}{}_{a} - 2 \hat R^{ab}{}_{;ab}$
which means that our analysis also includes the case $n=2$.},
 
$$ \eqalign{\hat R^{;a}{}_{a} - 2 \hat R^{ab}{}_{;ab}  & =  4 H^{ij}{}_{i;j}{}^a{}_a -4 (H^i{}_{a[i;b]}+
H^i{}_{b[i;a]})^{ab}\cr & =  2R_{,a}H^{aj}{}_j + 4 R^{ab}H_{a}{}^j{}_{j;b} - 4
R^{ab;c}H_{cab} + 2R^{abcd} H_{abc;d} . }
\eqno(7)$$
In flat space this equation is the triple trace of  (5).

The next step is the important one; by substituting  $\hat R_{abcd}$ in the last term  on the right hand
side we eliminate some of the remaining ackward potential terms as well as introducing the Riemann-candidate
explicitly also on the righthand side,
$$
\hat R^{;a}{}_{a} - 2 \hat R^{ ab}{}_{;ab}  = 2 R_{,a} H^{aj}{}_j+ 4R^{ab}H_a{}^j{}_{j;b} - 4
R_{ab;c}H^{cab}+ {1\over 2}R^{aijk} \hat R_{aijk}
\eqno(8)$$
By  decomposing $R_{abcd}$ and
$\hat R_{abcd}$ into their trace-free parts with $R_{ab}=S_{ab}+Rg_{ab}/n$, \  $\hat R_{ab}=\hat
S_{ab}+\hat R g_{ab}/n$, we obtain the alternative form,
$$
2\hat S^{ab}{}_{;ab} +{2-n \over n}\hat R^{;a}{}_a = ({4\over n}-2)R_{,a} H^{aj}{}_j-
4S^{ab}H_a{}^j{}_{j;b}+ 4 S_{ab;c}H^{cab}- {1\over2}C^{abcd}\hat C_{abcd}+{2\over n-2} S^{ab}\hat S_{ab} +
{n-2 \over n}R \hat R . \eqno(9)
$$
\underbar {$n=2$} 

Equation (9) is not valid for $n=2$, but
the previous equation (8) is. However, when we substitute 
$n=2$ with $R_{abcd}=Rg_{a[c}g_{d]b}$ and $\hat R_{abcd}=\hat R g_{a[c}g_{d]b}$ into (8) the constraint collapses
to a trivial identity, and so in two dimensions this particular constraint is not effective.

\underbar {$n >  2$} 

We cannot  conclude that (9) is an {\it effective} constraint on all geometries, and
on all Riemann candidates, because of the existence of the potential $H_{abc}$ and its derivatives alongside the 
Riemann candidate $\hat R_{abcd}$ and the Riemann tensor $R_{abcd}$. However for Einstein spaces, 
($S_{ab}=0=R_{,a}
$) we obtain an expression with no explicit terms in the potential $H_{abc}$,
$$
2\hat S^{ab}{}_{;ab} +{2-n \over n}\hat R^{;a}{}_a = - {1\over2}C^{abcd}\hat C_{abcd} +
{n-2 \over n}R \hat R  \eqno(10)
$$
and in particular, for spaces of constant curvature,  we obtain 
$$
2\hat S^{ab}{}_{;ab} +{2-n \over n}\hat R^{;a}{}_a = 
{n-2 \over n}R \hat R . \eqno(11)
$$
Therefore, for Einstein spaces, we find that the existence of a potential $H_{abc}$ in (3) leads to an
{\it effective} constraint because the terms involving the potential explicitly all disappear, and we get a
condition (10) directly linking the background space geometry via the Riemann tensor
$ R_{abcd}$ with the Riemann candidate
$\hat R_{abcd}$. 

There are some very special situations where this restriction (10) is satisfied trivially; e.g.,
if the Riemann candidate
$\hat R_{abcd}$ satisfied a Bianchi like equation, $\hat R_{ab[cd:e]}=0$ and also has its Weyl and Ricci scalar
part zero, $\hat C_{abcd}=0=\hat R$. Of course we cannot conclude that in such situations a Lanczos potential
will exist; we must also remember that there could be additional  constraints  at this order, or at
higher orders of differentiation.
\smallskip
Turning to spaces other than Einstein spaces; important questions, with respect to a particular non-Einstein
space, are if there exist more constraints, and whether all Riemann-candidates  or some Riemann candidates or
no Riemann candidates can be generated by a potential from (3). It is easy to see that there must always be
\underbar{some} Riemann candidates, since  in  a particular space, if we choose a particular tensor $H_{abc}$
with the symmetries (2), we can then
\underbar{define} a Riemann-candidate via (3) which will automatically satisfy the constraint (9). However, our
analysis is unable to tell us whether, in  all non-Einstein spaces, for
\underbar{arbitrary} Riemann-candidates there exists a potential $H_{abc}$ with the symmetries (2) such that both
the sytem (3) and the constraint (9) are satisfied. 

 Therefore what we are able to conclude for
the Riemann-candidate--Lanczos problem is that:

$\bullet$ for $n = 2$  the  constraint (9) linking the Riemann-candidate tensor and the geometry is trivially
satisfied,

$\bullet$ for $n > 2$  the  constraint (9) linking the Riemann-candidate tensor and the geometry is
effective in some spaces, e.g., (10),  and so \underbar{not all} Riemann-candidates can admit Lanczos potentials
in all spaces via the representation (3);

$\bullet$ for $n >2$,  we  know that there are \underbar{some} spaces and Riemann-candidate
tensors which can admit Lanczos potentials via the representation (3) with (9) also satisfied identically; 
we do not know if there are more situations where Riemann-candidate
tensors  can admit Lanczos potentials via the representation (3), but if there are, then (9) will be  satisfied
identically; we also do not know if there are more constraints.

\medskip

\underbar{\it Riemann-Lanczos problem in
$n$-dimensional spaces}

When we specialise to the Riemann-Lanczos problem, i.e., $\hat R_{abcd}\equiv  R_{abcd}$, we find
that, as a consequence of (1), the left hand side of (9) is  identically zero via the contracted Bianchi identity
 giving,
$$
0 = ({4\over n}-2)R_{,a} H^{aj}{}_j-
4S^{ab}H_a{}^j{}_{j;b}+ 4 S_{ab;c}H^{cab}- {1\over2}C^{abcd}C_{abcd}+{2\over n-2} S^{ab} S_{ab} +
{n-2 \over n}R^2  \eqno(12)
$$
We can then conclude that for
$n >2$ we cannot have a Lanczos potential $H_{abc}$ for a space of (nonzero) constant curvature, and for $n\ge
4$ the only Einstein spaces which can have a Lanczos potential are those subjected to the 
restriction, 
$$ C^{aijk} C_{aijk} =
2{n-2 \over n}R^2 .
\eqno(13)$$
This restriction is  {\it effective} since there are no explicit terms involving the potential, and what we have
is a direct condition on the geometry, which clearly not all Einstein spaces satisfy; in fact (13) is an
additional invariant condition linking two Riemann scalar invariants in Einstein spaces.

This restriction (12) is only one scalar equation and  so in general it, in itself, would not appear to be a very
strong restriction on the class of Riemann tensors. For instance, in four dimensions we know that there exists
fourteen  Riemann scalar invariants --- in general; however, when we specialise to vacuum
 4-dimensional spacetimes  we note  that this constraint (13)  excludes all Petrov types of the Weyl
tensor except the very specialised Petrov type N. So, in vacuum in 4-dimensional spacetimes, (13) is a very
strong restriction. Although Petrov type N spaces are not restricted by (13), we cannot conclude that they
admit Lanczos potentials via (1); we must again remember that there could be additional  constraints  at this
order, or at higher orders of differentiation.

As regards spaces other than Einstein spaces we know that there are a few explicit special examples of Riemann
tensors with Lanczos potentials, e.g., some conformally flat spaces given in [9], and Debever, G\"odel, Kasner
4-dimensional spacetimes and a G\"odel 3-dimensional space given in [1,2].

To confirm the significance of the integrability condition (12) for the Riemann--Lanczos problem we have shown
that all of these special examples (with the exception of the Kasner spacetime where the calculations were too
complicated) are non-Einstein spaces and we have demonstrated explicitly that they satisfy
(12)\footnote{${}^{\ddag}$}{In [8], three examples of conformally flat 4-dimensional spacetimes are given and it
is simple to confirm that the respective Lanczos potentials satisfy (12). In [2], Lanczos potentials are given
for  an example of a  Debever and  G\"odel 4-dimensional spacetimes, and in [1] a Lanczos potentials is given
for  an example of a  G\"odel 3-dimensional space; in these spaces it is straightforward, with the help of
{\it GRTensorII} [18], to confirm that (12) is satisfied.}.  Whether potentials can be found for the Riemann
tensors of
\underbar{all} non-Einstein spaces cannot be decided from the above analysis.

Therefore what we are able to
conclude for the Riemann-Lanczos problem is that:

$\bullet$ for $n = 2$  the  constraint (12) on the Riemann tensor  is trivially
satisfied.

$\bullet$ for $n > 2$  the  constraint (12) on the Riemann tensor is effective in some spaces, e.g.
(13),  and so 
\underbar{not all} Riemann tensors can admit Lanczos potentials via the representation (1);

$\bullet$ for $n > 2$,  we do  know that there are \underbar{some} special examples of Riemann tensors which can
admit Lanczos potentials via the representation (1), and they  also satisfy (12); we do not know if there are any
others, but if there are, then the constraint (12) must be satisfied; we do not know if there are any more
constraints.

\medskip
 
The existence of this constraint for the Riemann--Lanczos problem was originally demonstrated in four
dimensions by Massa and  Pagani [5] who set up  the problem  in ordinary tensor
notation, but carried
out the actual derivation of the crucial constraint equation   in  tensor-valued
differential forms; this calculation was  quite involved, and strictly  
4-dimensional\footnote{${}^{\dag}$}{In [5] a different sign convention was used for the Ricci tensor
from that used in this paper, and this convention was also used in [6,7];  so there are some sign differences
between this version of the equation and that in [5,6,7].}.   The tensor-valued differential form
part of the derivation of the integrability condition was  rederived by Edgar in [6] in ordinary tensor notation,
but the argument was still strictly  4-dimensional. Subsequently, a more  direct and complete
derivation of the constraint equation --- with no explicit dimension imposed --- was given in [7], and an
even simpler variation by H\"oglund [19]. The derivation given above for  the
Riemann-candidate--Lanczos problem is based on the version in [19].

\

{\bf 3. Effective Constraints  for the Parallel Problem.}

In their investigations Bampi and
Caviglia [8,9] did not in fact deal with $\hat R_{abcd}$ and $H_{abc}$ directly but rather with their respective
counterparts $N_{abcd}$ and $T_{abc}$ which satisfied 
$$N_{abcd} = \  2 T_{ab[c;d]}
+2T_{cd[a;b]}   \eqno(14)$$

where $N_{abcd}$ and $T_{abc}$ have only the respective symmetries,
 $$T_{abc}=T_{[ab]c}\qquad\qquad \hbox{and}
\qquad\qquad N_{abcd}=N_{[ab][cd]}=N_{cdab} .  \eqno(15)$$  
Their motivation for studying this parallel problem was that they were able to show that 
 this problem and the
Riemann-candidate--Lanczos problem were mathematically equivalent --- {\it in four dimensions}. For other
dimensions, any positive results for the existence of potentials for all $N_{abcd}$ would also apply to the
narrower  Riemann-candidate--Lanczos problem; but negative results for the parallel problem would, in general,  be
irrelevant to the
narrower  Riemann-candidate--Lanczos problem.
\medskip
\underbar{$n > 4$.} 

We can immediately find the integrability condition 
$$N_{[abcd;e]} = 0 \eqno(16)
$$
and confirm that this is always an {\it effective} constraint. So we can conclude that in this parallel problem
not all tensors $N_{abcd}$ can be written in terms of a potential; this result does not permit us to draw any
conclusion about the associated Riemann-candidate--Lanczos problem.
\medskip
 \underbar{$n > 2$.}

If we carry out again the antisymmetrisation over 5 indices as in (16) we just obtain the trivial identity in
 dimensions 3 and 4.  So instead, to find effective constraints for $n > 2$, we have to carry out the same
procedure as in Section 2 involving two differentiations; but since we already know that there are restrictive
integrability conditions for the Riemann-candidate--Lanczos problem, there is no purpose in investigating
further the parallel problem as a means of investigating the narrower Riemann-candidate--Lanczos problem.
However, for completeness we add that in the calculations leading to the constraint (8), the only index
symmetries used were those of the type (15), and so we can deduce that the parallel problem is subject to the
constraint
$$
N^{;a}{}_{a} - 2 N^{ ab}{}_{;ab}   = 2 R_{,a} T^{aj}{}_j+ 4R^{ab}T_a{}^j{}_{j;b} - 4
R_{ab;c}T^{cab}+ {1\over 2}R^{aijk} N_{aijk} .
\eqno(17)$$
which in flat space simplifies to
$$
N^{;a}{}_{a} - 2 N^{ ab}{}_{;ab}  = 0 .
\eqno(18)$$
As argued in the last section, these constraints are effective.
\medskip

In summary, we can conclude that, since our investigation found the existence of effective  constraints in the
parallel problem, we cannot draw any conclusions about the original Riemann-candidate--Lanczos
problem\footnote{${}^{\ddag}$}{Of course if the results for the parallel problem had been positive, then these
results could have been  specialised to the narrower Riemann-candidate--Lanczos problem. This is the case in the
related Weyl-candidate--Lanczos problem also considered in [9].}, since in all cases  the constraints in the
parallel problem may or may not be present in the narrower Riemann-candidate--Lanczos problem.  However, the
occurence of the integrability conditions (16) and (17,18), and the role played by dimension will be of interest
in the next section.

\

{\bf 4. 'Generic', 'Ordinary' and 'Singular' solutions.}

From the type of investigaton   in Sections 2 and 3 on the
constraints due to integrability conditions, we are only able  to directly draw limited conclusions. For more
complete conclusions we need a procedure which will distinguish between the respective situations where there
are indeed  effective constraints (even if we cannot find them explicitly), and where there 
are no effective constraints and the existence of a potential is always
guaranteed.  This is clearly the role of Cartan's local criteria of integrability of ideals of exterior forms
[20,21,22], as set out in [8,9]. Furthermore, if the system is not in involution and we do find some effective
constraints, we can prolong the original system to take account of these constraints, and then also use Cauchy's
criteria to analyse the prolonged system; if the prolonged system is not in involution then the process can be
repeated.

As noted in the last section  Bampi and
Caviglia [8] considered the parallel problem (14)  for  the  tensors $N_{abcd}$ and
$T_{abc}$ as a means of studying the 
Riemann-candidate--Lanczos  problem (3). We shall now discuss their results, and compare with our results
in the previous two sections.  

\underbar {$n > 4$.} In [8] it is  stated that in higher dimensions there will be non-trivial restrictions on
the data. In fact we have obtained this result  very easily in (16).  This result has no direct relevance
to the narrower Riemann-candidate--Lanczos problem.

\underbar{$n=4$ }

In their first paper, Bampi and
Caviglia [9]   showed that the equation (14)  does not always admit a solution for a given $
N_{abcd}$. 
More precisely, they showed the non-existence of solutions under certain generic
conditions on $
N_{abcd}$, or as Massa and Pagani [5] pointed out, Bampi and
Caviglia [9] showed that the
representation (3)  does not exhaust the totality of  the set of tensors $N_{abcd}$
(which in four dimensions is equivalent to the set of Riemann-candidates $
\hat R_{abcd}$), and  
 hence, since the Riemann tensors themselves are only a proper 
subset of this
larger class of Riemann-candidates, this result says nothing about the validity of (3)  for 
Riemann tensors.

 This  generic result for the Riemann-candidate--Lanczos problem given by  Bampi and
Caviglia [9] was strengthened in their second paper [8] where it was
stated in Theorem 1  that there never exists 'regular' ('ordinary')  solutions to (14) for any
tensor $N_{abcd}$ in four dimensions; so this result now includes the Riemann--Lanczos problem, i.e., there
never  exists 'regular' ('ordinary')  solutions to (1) for any Riemann
tensor $R_{abcd}$ in four dimensions.
If we interpret 'regular' solutions to mean the existence of the most general solutions 
with no constraints on
the class of tensors
$N_{abcd}$ and the underlying space, then the conclusion in Theorem 1 does not contradict the results in Section
3 of this paper.

The difficulty is with Theorem 2 in [8]. The original system (14) is not in involution and so is prolonged,
and as a result of the  analysis of the
prolonged system, Bampi and
Caviglia find no  constraints, and conclude in  Theorem 2 [8]  that in four
dimensions, although the representation (14) never permits 'regular' ('ordinary')  solutions to (14), it
 \underbar{always} admits 'singular' ('nonordinary') solutions\footnote{${}^{\dag}$}{There is no precise
explanation in [8] of the difference between   'regular' and 'singular' solutions, but rather an analogy is
given. It would seem that Bampi and Caviglia [8]  view the distinction as simply a technical
matter: solutions involving the maximal Cartan characters are 'regular' solutions, while solutions from a
prolonged system with less than the maximal Cartan characters are  'singular' solutions.} for all tensors
$N_{abcd}$ (equivalently for all Riemann-candidates
$
\hat R_{abcd}$).  Therefore, in
[8] the 'singular' solutions are argued to have what appears to be exactly the same general properties as
'regular' solutions, with no constraints on the data $N_{abcd}$ or on the background space. 

Normally we expect that such 'singular' solutions would involve some
restrictions on the set of tensors $N_{abcd}$ and/or on the underlying space. 
However, Bampi and
Caviglia [8] claim that these
'singular' solutions are, to their own surprise, not subject to any integrability conditions, and therefore 
{\it the class of
'singular' solutions allows a potential to exist without any restriction
on $N_{abcd}$ and on the geometric structure of the underlying Riemann manifold}.

This conclusion in Theorem 2 [8]  contradicts the results in
Section 3 of this paper; in
particular the very simple obvious effective constraint (17) for flat space (flat space is not excluded in the
analysis in [8]). 
 We suspect that this claim of no constraints on
these 'singular' solutions is  not due simply to a misinterpretation of the properties of 'singular' solutions
but, more fundamentally, to a fault in the calculations in [8].  We would point out that   the only
constraint --- the 'internal identity'
$A_{[abcde]}=0$   in the notation of [8] --- discovered in the calculations for Theorem 2 in [8]
is precisely the very obvious integrability condition (18) which we found directly in Section 3, and which is
of course trivially satisfied in four dimensions.   What is most
surprising about the  method of application of Cauchy's criteria in [8,9] is that the possibility of
 constraints existing after two differentiations --- involving linear combination of components of 
$B_{abcdef}$ in the notation of [8,9] --- does not arise; whereas, from our work in Section 3, and in
particular the simple effective constraint equation (18), it is clear that this is precisely where we expect  
constraints. Unfortunately, it is not easy to check the accuracy of  the argument in [8] at this level,
since no explicit details were given leading to the conclusion that
$s'_3=s_3$, for the Cartan characters.  

\medskip

\underbar {$n=3$.} For the parallel problem, it is  stated in [8] that, under 'generic' conditions,  there does
not exist any 'regular' solution because of the existence of an internal identity; however in their second paper
[8] Bampi and
Caviglia   state that, by the same argument as in four dimensions, there exist 'singular' solutions independently of the
choice of
$\hat R_{abcd}$ and the geometry of the space.   So we have here the same situation as in four dimensions, involving a
contradiction with the  effective constraint  found in Section 3.  
Of course, we should remember that the parallel problem is not equivalent to the Riemann-candidate--Lanczos
problem in three dimensions, and a negative result in the former has no direct relevance to the latter.

Significantly, in the next section, we will find that Dolan and Gerber [1], using an alternative but related
method to [8] for the direct Riemann--Lanczos problem, have found an explicit  constraint which is
effective; and this constraint is precisely the three dimensional version of the effective constraint which we
discussed in Section 2.

\medskip

\underbar{$n=2$.}  In [9] it is  stated that there will always be 'regular' solutions, and there is no
contradiction with the integrability condition (9) in Section 2 since with the substitution $n = 2$ the
constraint is trivially satisfied. 
\medskip

Finally we turn to the results on the differential gauge.
It is stated in [8] that even when an arbitrary differential gauge
condition on
$H_{ab}{}^c{}_{;c}$ (e.g. $H_{ab}{}^c{}_{;c}=0 $) is put alongside the condition (3), there will
always be 'singular' solutions in four dimensions.  As we have discussed above, we believe that the result
that there are
always 'singular' solutions is flawed.  But the question arises, in those special cases where
there are 'singular' solutions (with  a restriction on the Riemann-candidates and/or the geometry) whether the
differential gauge can be chosen arbitrarily.  We believe that this question of gauge is still, in general, open.

\

{\bf 5. The Riemann-Lanczos problem by the Janet-Riquier approach.}

Dolan and Gerber [1,2] have considered the direct Riemann--Lanczos problem (1) with the symmetries (2)
(not the parallel problem
as in [8,9]) from the exterior derivative viewpoint along the same lines as in [8,9]; but also they have
considered the problem directly as a system of partial differential equations in two and three dimensions [1]
using the related Janet-Riquier [23,24] approach. In fact their analysis is  valid also for the more
general Riemann-candidate--Lanczos problem (3).

\underbar{$n=2$.}  This  problem  has been shown by the Janet-Riquier approach always to
have solutions;  in fact it is a very simple problem which has also been integrated directly in [1].  There is
no contradiction with the integrability condition (9) in Section 2, since we have already noted for  $n=2$
that the constraint is not effective.

\underbar{$n=3$.}
Using the Janet-Riquier approach, Dolan and Gerber [1] have found that the original Riemann--Lanczos
problem is not in involution and so  there are no
'regular' solutions.  After one prolongation, obtained by adding  one 'internal identity', they found that the
prolonged system was involutive. 
They point out  that their 'internal identity' is not trivial, and they give it in invariant form as
$$
f^{(R)}{}_{12[12;3]3}+f^{(R)}{}_{23[12;3]1}+f^{(R)}{}_{31[12;3]2}=0 \eqno(19)
$$
where 
$$f^{(R)}{}_{abcd} \equiv R_{abcd} - 2 H_{ab[c;d]}
-2H_{cd[a;b]}\eqno(20) .
$$
 This is precisely the constraint
$$f^{(R)}{}_{[ab}{}_{|[de}{}_{;f]|}{}_{c]} =0 .\eqno(21)
$$
since there is only one component of (21) in three dimensions; and  since in three dimensions there is no 
trace-free part to
$f^{(R)}{}_{abcd}$, the constraint (21) can be rewritten more compactly as
$$2f^{(R)}{}^{aib}{}_{i;ab} - f^{(R)}{}^{ij}{}_{ij}{}^{;a}{}_{a}=0\eqno(22)
$$
When the substitution (20) is made into this last  equation, we obtain
$$  R^{;a}{}_{a} - 2  R^{ab}{}_{;ab}   =  4 H^{ij}{}_{i;j}{}^a{}_a -4 (H^i{}_{a[i;b]}+
H^i{}_{b[i;a]})^{ab}  .
\eqno(23)$$
which is precisely  (the Riemann tensor version of) equation (7), and leads to an effective constraint (8) as
shown. Of course, for  a Riemann tensor, the left hand side will be
indentically zero from  the Binchi identity, but we can see that the analysis also gives the constraint for the
more general Riemann-candidate--Lanczos problem. 

 {\it Therefore,  the 'internal identity' above (19)
  is precisely the three dimensional version of the effective constraint (12) for the Riemann--Lanczos
problem found in Section 2.} Furthermore, since the prolonged system, created by adding this constraint, has been
shown to be involutive in [1] this must be the only constraint.

Also in [1] there is  an explicit example in three dimensions of a Lanczos potential, which can be interpreted as
a singular solution for the unprolonged problem, or as a regular solution for the prolonged problem.  We have
confirmed directly that it satisfies the constraint (9) in Section 2 (equivalently the 'internal identity' (19)
which was found in [1]).

\

{\bf 6. Conclusion.}

We have confirmed that   two successive differentiations of the defining equations (3) of the Riemann-candidate
--Lanczos problem, leads to an effective constraint in $n > 2$ dimensions, and known solutions of the problem have been
shown explicitly to satisfy this constraint. Furthermore, we have shown that the results in [5,6,7] are very relevant
to the work of Dolan and Gerber [1,2]; in particular the 'internal identity' found in [1] for the
three dimensional Riemann--Lanczos problem is precisely the effective constraint found in [5,6,7].
The existence of this effective constraint  contradicts  the results in [8] for three and four
dimensions.   

It is significant that, using a method similar to the approach in [8,9],  in three dimensions, Dolan and
Gerber have found in [1] exactly the three dimensional version of the  effective constraint  originally found by
[5,6,7]. This 
reinforces our suspicion that the  approach in [8,9] is flawed, and an attempt should be made to reinvestigate the
prolonged system at the  level of the second derivatives of the Riemann tensor for all dimensions
from the exterior differential system viewpoint.  (Dolan and Gerber [2] have briefly discussed the problem in some other
dimensions from the exterior differential system viewpoint as used in [8,9], but have not developed it further.) 
However, it
seems that the Janet-Riquier method for partial differential equations is shorter and perhaps more transparent, and so
it would be preferable to first apply this approach, as used in two and three dimensions  in [1], to other
dimensions; then this could be compared with the exterior differential system approach.

We also note the significant role that the dimension of the space has played at a number of crucial places in
our arguments;  whether a constraint is effective or not can depend on the dimension. The constraint (8) is
effective for $n>2$, but trivial for $n=2$; in the parallel problem as well as the constraint (17) for $n>2$,
an additional constraint (16) occurs for $n>4$. The role of dimension is even more subtle for the Weyl-Lanczos
problem where complicated constraints involving the second derivatives of the Weyl tensor  are valid only for
dimensions
$n\ge 6$, while constraints involving the third derivatives of the Weyl tensor  are valid
only for dimensions
$n\ge 5$; there are no constraints for $n = 4$ [10 - 15].
\medskip
In summary, we note that not all Riemann candidates and Riemann tensors have potentials in dimensions $n> 2$,
because of the existence of an effective constraint; although there are special cases where such potentials do exist.
From [1] it is known that prolongation with this one constraint gives an involutive system, in three dimensions;  it is
still an open question whether prolongation with this one constraint can lead to involution in higher
dimensions; a preliminary investigation  of the flat space case by the Janet-Riquier approach would suggest
that 
\underbar{all} the equations corresponding to (5) may be needed.

\

\

{\bf Acknowledgements}

The continuing financial support of Vetenskapsr\aa det (the Swedish Research Council) is acknowledged.
Thanks to  Anders H\"oglund  and Fredrik Andersson for many helpful discussions.

\

\

{\bf REFERENCES}.

1. Dolan, P and Gerber, A. (2002). {\it Janet-Riquier Theory and the Riemann--Lanczos Problems in
2 and 3 Dimensions.} Preprint:   gr-qc/0212055

2. Dolan, P and Gerber, A. (2002). {\it The Riemann--Lanczos Problem as an Exterior Differential 
System with Examples in 4 and 5 Dimensions.} Preprint:  gr-qc/0212054

3. Brinis Udeschini, E. (1977).
 {\it Mech. Fis. Mat. Istituto Lombardo (rend. Sc.)}, {\bf 111}, 466. 
\ (1980) \ 
{\it Gen. Rel. Grav.}, {\bf 12}, 429.

4. Penrose, R.  and  Rindler, W. (1984). {\it Spinors 
and Spacetime Vols.1 and 2} (Cambridge University 
Press) .

5. Massa, E. and Pagani, E. (1984) \ 
{\it Gen. Rel. Grav.}, {\bf 16}, 805.

6. Edgar, S. B. (1987).  {\it Gen. Rel. 
Grav.}, {\bf 19},  1149.

7. Edgar, S. B. (1994).  {\it Gen. Rel. 
Grav.}, {\bf 26},  329.

8. Bampi, F., and Caviglia, G. (1984). {\it Gen. Rel. 
Grav.}, {\bf 16}, 423.

9. Bampi, F., and Caviglia, G. (1983). {\it Gen. Rel. 
Grav.}, {\bf 15}, 375.

10. Lanczos, C. (1962). {\it  Rev. Mod. Phys.}, {\bf 
34}, 379.

11. Edgar, S. B. and H\"oglund A. (1997) {\it Proc. Roy. Soc. A}, {\bf 453}, 835.

12. Illge, R. (1988). {\it Gen. Rel. Grav.}, {\bf 20}, 
551.

13. Andersson, F. and Edgar S.B. (2001).   {\it Class. Quantum Gravity}, {\bf 18}, 2297.

14. Edgar, S.B. and H\"oglund A. (2000). 
 {\it Gen. Rel. Grav.}, {\bf 32},  2307.

15. Edgar, S.B. and H\"oglund A. (2002).  {\it Gen. Rel. Grav.}, {\bf 34}, 2149.

16.  Lovelock, D. (1970). {\it Proc. Camb. Phil. Soc.}, 
{\bf 68,} 345.

17.  Edgar, S.B. and H\"oglund A. (2002).  {\it  J. Math. Phys.} {\bf 43}, 659.

18. Available at http://grtensor.org

19. H\"oglund, A. (1955). {\it The Lanczos Potential and its Wave Equation}. M.Sc. dissertation,
LiTH-MAT-Ex-95-08. Department of Mathematics, Link\"{o}pings universitet, Sweden.

20. Cartan, E.  {\it Les syst{\'e}m{\`e}s diff{\'e}rentiels ext{\'e}rieurs et leurs
  applications g{\'e}om{\'e}triques.} Hermann,
Paris, (1971).

21. Hermann, R. (1965). {\it Adv. Math.}, {\bf 1}, 265.

22. Kuranishi, M.  {\it  Lectures on Involutive Systems of Partial Differential
  Equations.}  Publica\c{c}\~{o}es da Sociedade de Mathem\'{a}tica de S\~{a}o Paulo,
  S\~{a}o Paulo, (1967).

23. Janet, M.  (1920). {\it J. Math. pures et appl.}, {\bf 3}, 65. 

24. Riquier, C. 
 {\it Les syst\`{e}mes d'\'{e}quations aux deriv\'{e}es partielles.}
 Gauthier-Villars, Paris, (1910).

\end